\documentclass{article}
\usepackage[utf8]{inputenc}
\usepackage{amsmath}
\usepackage{amssymb}
\usepackage{amsfonts}
\usepackage{esvect}
\usepackage{cite}
\usepackage{physics}
\usepackage{mathrsfs}
\usepackage{authblk}
\usepackage{geometry}
\usepackage{abstract}
\usepackage{xcolor}
\title{\bf Hidden symmetry algebra and construction of quadratic algebras of superintegrable systems }
\author{\large Rutwig Campoamor-Stursberg$^{1}$\footnote{rutwig@ucm.es}, 
Ian Marquette$^{2}$ \footnote{i.marquette@uq.edu.au} }
\affil{$^{1}$ Instituto de Matem\'atica Interdisciplinar and Dpto. Geometr\'{\i}a y
Topolog\'{\i}a,
UCM,\\E-28040 Madrid, Spain}
\affil{$^{2}$ School of Mathematics and Physics, The University of Queensland \\ Brisbane, QLD 4072, Australia}

\usepackage[colorlinks=true,urlcolor=black]{hyperref}

\newcommand{\g}{\mathfrak{g}}
\newcommand{\gl}{\mathfrak{gl}}
\newcommand{\RR}{\mathbb{R}}

\begin{document}

\maketitle
\begin{abstract}
The notion of hidden symmetry algebra used in the context of exactly solvable systems (typically a non semisimple Lie algebra)  is re-examined from the purely algebraic way, analyzing subspaces of commuting polynomials that generate finite-dimensional quadratic algebras. By construction, these algebras do not depend on the choice of realizations by vector fields of the underlying Lie algebra, allowing to propose  a new approach  to analyze polynomial algebras as those subspaces in an enveloping algebra that commute with a given algebraic Hamiltonian.  These polynomial algebras play an important role in context of superintegrability, but are still poorly understood from an algebraic point of view. Among the main results, we present polynomial quadratic algebras of dimensions $4, 5, 6$ and $8$, as well as cubic algebras of dimensions $3$ and $5$, and various Abelian algebras, all of dimension $3$. Basing on the observation how superintegrability is associated with exact solvability, we propose a procedure that connects the underlying Lie algebra with algebraic integrals of motion. As the integrals constructed in such way are now independent on the realization, alternative choices of realizations  can provide new explicit models with the same symmetry algebra. In this paper, we consider examples of such equivalent Hamiltonians in terms of differential operators for the three cases and connected to the underlying Lie algebra $\gl(2,\mathbb{R}) \ltimes \RR^{2} \oplus T_{1}$ as well as to the maximal parabolic subalgebra of $\mathfrak{gl}(3,\mathbb{R})$. We also point out differences between the enveloping algebra of Lie algebras and the enveloping algebra of the related differential operators realization.  
\end{abstract}

\section{Introduction}

The idea of hidden symmetry Lie algebra, introduced in regard of exactly and quasi-exactly solvable (quantum) models is based on the existence of an infinite flag of functional linear spaces that is preserved that the Hamiltonian,  such as e.g. the
finite-dimensional representation spaces of (semi-simple) Lie algebras of
first order differential operators \cite{tur88,tur89,Tur92}.  This allows to describe the Hamiltonian of a system in terms of a parabolic subalgebra of a Lie algebra. One the most relevant features that such an approach  offers is the possibility to obtain the spectrum of quantum models using tools from representation theory \cite{tur94,tem01}. However, this scheme is somewhat limited, as it depends heavily on explicit realizations of Lie algebras by means of differential operators. Other classes of quantum models and superintegrable systems have been studied over the years using different techniques, such as systems characterized by symmetry algebras related to non Abelian polynomial algebras (see e.g. \cite{das10,mil13,mil14,hoq15,liao18, chen19}). The underlying polynomial algebras are constructed via integrals based on explicit differential operator realizations \cite{chen19}, a fact that again poses a more or less severe restriction  for their classification and detailed study, as the relations among elements must be understood via a differential operator algebra.

\smallskip

The purpose of this paper is to describe, through the analysis of some representative quasi-exactly solvable systems, how purely algebraic polynomial algebras can be constructed in such manner that the integrals deduced also have an algebraic origin. The scheme is based on the connection between hidden symmetries and symmetry algebras \cite{tem01,tre09}. In these earlier works, the symmetry algebra is a consequence of the explicit choice of realization. In this work we reexamine these examples and construct polynomials in terms of the generators of a parabolic subalgebra that commute with the Hamiltonian, which is a second-order polynomial in the generators. Only the commutation relations of the underlying non semisimple Lie algebra are used. It turns out that these polynomials span the space of polynomials which commute with the Hamiltonian and provide a finitely-generated polynomial algebra of higher order. As the ansatz is purely algebraic, this points out the freedom in choosing the explicit realization of the Lie algebra underlying the exact solvability of a system. Using the realization related to the coadjoint representation, we present three examples exhibiting differential operators that can be defined as the Hamiltonian of the system,  and allow the existence of a quadratic algebra equivalent to the starting one, hence leading to an algebraic equivalence. To our authors's knowledge,  such an approach has not been considered previously in the literature.

\subsection{Finite-dimensional quadratic and hidden symmetry algebras}

\noindent One-dimensional and spherically-symmetric quasi-exact solvable problems have been extensively studied, and shown to be related to the representation theory of the simple Lie algebra $\mathfrak{sl}(2,\mathbb{K})$ with $\mathbb{K}=\mathbb{R},\mathbb{C}$ (see \cite{tur88} and references therein). More specifically, realizing $\mathfrak{sl}(2,\mathbb{R})$ by the vector fields
\begin{equation}
J^+_n = x^2 \partial_x - n x\ \ ,\ \ J^0_n = x \partial_x - {\frac{n }{2}}\ \ ,\ \ J^-_n =
\partial_x, \quad n\in\mathbb{N},  \label{sl2}
\end{equation}
the space  $\mathcal{P}_n=\langle x^k|0\leq k \leq n \rangle$ of polynomials of degree at most $n$ is easily seen to be invariant by the action of the operators in  (\ref{sl2}). Considering $n$ as a parameter, we obtain the flag space 
\begin{equation}
\mathcal{P}_0\subset \mathcal{P}_1\subset \dots \subset\mathcal{P}_n\subset \mathcal{P}_{n+1}\subset \dots, \label{flag0}
\end{equation}
that turns out to be preserved for any value of $n$ by a generic element of the universal enveloping algebra $\mathcal{U}(\mathfrak{b})$
of the Borel subalgebra $\mathfrak{b}$ of $\mathfrak{sl}(2,\mathbb{R})$ generated by the operators $J^0_n, J^-_n$. \textit{Exactly solvable} systems are obtained considering the Hamiltonian of the system as an element of the  enveloping algebra $\mathcal{U}(\mathfrak{b})$, while quasi-exactly solvable systems are defined requiring that the Hamiltonian
leaves the subspace $\mathcal{P}_n$ invariant. The latter algebraic condition actually characterizes quasi-exactly solvable one-dimensional Hamiltonian systems, whenever the generators are  taken in
realization (\ref{sl2}), as was shown in \cite{Tur92}.  
\noindent The natural generalization of the one-dimensional case is obtained considering the $N$-dimensional analogue of the flag space (\ref{flag0}) identified with finite-dimensional representations of semi-simple Lie algebra $\mathfrak{g}$ realized as 
first-order differential operators, and imposing that the Hamiltonian of the system is expressed in terms of the enveloping algebra associated to a distinguished subalgebra that preserves the flag, such as a parabolic subalgebra (see e.g. \cite{tem01}). 
The algebraic formalism underlying the method of the hidden algebra can be briefly summarized in the following steps: 

\medskip
\noindent Let $\mathfrak{g}$ be a (semisimple) Lie algebra and let $\Phi:\mathfrak{g}\rightarrow \frak{X}(\RR^2)$ be a realization of the Lie algebra by first-order differential operators such that  
for any $n$, the  linear space of polynomials of degree $p\leq n$ defined by   
\begin{equation}
\mathcal{P}^{2}_n=\left\{ t^{k}u^{m},\quad 0\leq k+m\leq n \right\}\label{flag}
\end{equation}
is a finite-dimensional representation of $\mathfrak{g}$. Considering $n$ as a parameter, it follows that the flag space 
\begin{equation}
\mathcal{P}^{2}_0\subset \mathcal{P}^{2}_1\subset \dots \subset\mathcal{P}^{2}_n\subset \mathcal{P}^{2}_{n+1}\subset \dots \label{flag2}
\end{equation}
is invariant by the action of $\Phi(\mathfrak{g})$. Now suppose that $\mathfrak{m}$ is a subalgebra of $\mathfrak{g}$ such that for each $n\geq 0$ the relation
\begin{equation}
 X\left(P^2_n\right)\subset P^2_n,\quad \forall X\in\Phi(\mathfrak{m})\label{flag3}
\end{equation}
holds, i.e., that the flag is preserved by the (realized) subalgebra,\footnote{The case where $\mathfrak{g}\simeq \mathfrak{m}$ is not excluded.} and such that the Hamiltonian $h$ of the system can be expressed in terms of the differential operators associated to the generators in $\mathfrak{m}$: 
\begin{equation}
h= \sum_{i,j=1}^{\dim\frak{m}} \alpha_{ij}\Phi(X_i)\Phi(X_j)+\sum_{k=1}^{\dim\frak{m}} \beta_k\Phi(X_k),\label{ham1}
\end{equation}
where $\alpha_{ij},\beta_{k}$ are constants and $\left\{X_1,\dots ,X_{\dim\frak{m}}\right\}$ is a basis of $\mathfrak{m}$. In this context, the Hamiltonian $h$ can be interpreted as the image, via the realization $\Phi$, of a quadratic element $H$ in the universal enveloping algebra $\mathcal{U}(\mathfrak{m})$. Similarly, the constants of the motion $\varphi_1,\varphi_2$ can also be rewritten as elements of the enveloping algebra. As differential operators (i.e., evaluated in the realization $\Phi$), they satisfy the commutators  
\begin{equation}
\left[ h,\varphi_1\right]= \left[ h,\varphi_2\right]=0.\label{flag5}
\end{equation}
The commutator $\left[ \varphi_1,\varphi_2\right]$, as well as successive commutators provide additional (dependent) higher-order constants of the motion. It follows that, in general, the operators $h,\varphi_1,\varphi_2$ generate an infinite-dimensional algebra. 

\medskip
\noindent The use of hidden symmetry algebras in the context of (quasi-)exact solvable systems may suggest a purely algebraic procedure to construct finite-dimensional  quadratic polynomial algebras. In the preceding conditions, we can formally consider the polynomial
\begin{equation*}
H= \sum_{i,j=1}^{\dim\frak{m}} \alpha_{ij} X_i X_j +\sum_{k=1}^{\dim\frak{m}} \beta_k X_k.\label{hafu}
\end{equation*}
This element of $\mathcal{U}(\mathfrak{m})$ corresponds to the Hamiltonian $h$ once the generators of $\mathfrak{m}$ are realized by vector fields. Now let $J_1,J_2$ be two polynomials such that they correspond, via the realization $\Phi$, to the constants of the motion $\varphi_1,\varphi_2$. From the purely algebraic point of view, there is no necessity that the elements $J_s$ commute with $H$ as polynomials in $\mathcal{U}(\mathfrak{m})$, as we merely know that the identity 
\begin{equation}
\left[h,\varphi_s\right] =  \left[ H,J_s\right]=0 \quad ({\rm mod}\quad \Phi)\label{rax1}
\end{equation}
holds. Hence, the operators $\varphi_s$ are first integrals of the Hamiltonian $h$ as a consequence of the specific realization $\Phi$, and not because  they commute in the enveloping algebra  $\mathcal{U}(\mathfrak{m})$. 

\medskip
\noindent The question that arises naturally in this context is whether we can find polynomials in the enveloping algebra $\mathcal{U}(\mathfrak{m})$ that commute with $H$ and generate a finite-dimensional quadratic Lie algebra, independently of any particular realization. Starting from $H\in\mathcal{U}(\mathfrak{m})$ and $d= 2$, let $\frak{M}_2$ denote the set of quadratic elements in the enveloping algebra such that they commute with $H$:
\begin{equation*}
\frak{M}_2=\left\{ P\in\mathcal{U}(\mathfrak{m})\;\left |\right.\; \left[H, P\right]=0,\; \frac{\partial^{3}P}{\partial X_{i_1}\partial X_{i_2}\partial X_{i_{3}}}=0,\; 1\leq i_1\leq i_2\leq i_3\leq \dim\frak{m}\right\}.
\end{equation*}
The set $\frak{M}_2$ in particular contains the Hamiltonian $H$. If $\frak{J}_2=\left\{I_1^{(2)},\dots ,I_q^{(2)}\right\}$ are linearly independent elements of $\frak{M}_2$, there certainly exist $ n_j\in\RR$ such that 
\begin{equation*}
H +\sum_{j=1}^{q} n_{j} I_{j}^{(2)}=0\quad .
\end{equation*}
holds. Along the same lines, for any order $d > 2$, let $\frak{M}_d$ be the set of elements in the enveloping algebra of degree not exceeding $d$ such that they commute with $H$:
\begin{equation*}
\frak{M}_d=\left\{ P\in\mathcal{U}(\mathfrak{m})\;\left |\right.\; \left[H, P\right]=0,\; \frac{\partial^{d+1}P}{\partial X_{i_1}\dots\partial X_{i_{d+1}}}=0,\; 1\leq i_1\leq\dots \leq i_{d+1}\leq \dim\frak{m}\right\}.
\end{equation*}
For each such index, we denote by $\frak{J}_d$ the set of linearly independent elements. By construction, we have the  filtration $\frak{M}_2\subset \frak{M}_3\subset \dots \subset \frak{M}_d\subset \dots$ within the centralizer $C_{H}\mathcal{U}(\frak{m})$ of $H$ in the enveloping algebra \cite{Dix}. In particular, the inclusions  $\frak{J}_2\subset \frak{J}_3\subset \dots \subset \frak{J}_d\subset \dots$ hold. For the commutators of elements we have the inclusion 
\begin{equation}
\left[\frak{M}_p,\frak{M}_q\right]\subset \frak{M}_{p+q-1},\quad p,q,\geq 2
\end{equation}
is satisfied. 

\medskip
\noindent Fixing a value $d_0\geq 2$, a subset $\frak{N}_{d_0}=\left\{P_1,\dots ,P_r\right\}\subset \frak{J}_{d_0}$ generates a finite-dimensional quadratic polynomial algebra $\frak{A}\left(\frak{N}_{d_0}\right)$ if for any $1\leq i,j\leq r$ there exist  constants $\mu_{ij}^{k\ell}, \nu^{q}_{ij}$ such that 
\begin{equation}
\left[ P_i,P_j\right]= \sum_{k,\ell=1}^{r} \mu_{ij}^{k\ell}P_kP_\ell+\sum_{q=1}^{r} \nu_{ij}^{q}P_q.\label{quad1}
\end{equation}

\medskip
\noindent Therefore, given a (quasi-)exactly solvable system with Hamiltonian $h$, we can proceed recursively analyzing the commutators in $\mathcal{U}(\mathfrak{b})$ of the operators in $\frak{J}_d$ for any $d\geq 2$.

\section{Polynomial algebras related to the Smorodinsky--Winternitz system}

In the paper \cite{tre09},  the authors analysed the infinite-dimensional finitely-generated Lie algebra parameterized by an index $k$ (denoted with $s$ in \cite{tre09})
\[ \g^{(k)} \supset \gl(2,\mathbb{R}) \ltimes \RR^{k+1} \oplus T_{k} \]
using the following realization (see equations (19) and (20) there): 
\begin{equation}
 J_{N}^{1}=\partial_{t} ,\quad J_{N}^{2}= t \partial_{t}-\frac{N}{3},\quad J_{N}^{3}=ku \partial_{u} -\frac{N}{3},\quad J_{N}^{4}=t^{2}\partial_{t} +ktu \partial_{u}- Nt,\\
 R_{i}=t^{i}\partial_{u},\quad    T_{k}=u \partial_{t}^{k},\label{real}
\end{equation}
where $i=0,1,\dots,k$ and $k$ takes integer values. For the special values $k\leq 1$, we get, among others, the following commutators:

\begin{equation}
\begin{array}[c]{llll}
[J^{1},J^{2}]=J^{1}, & [J^{1},J^{3}]=0, & [J^{1},J^{4}]=2J^{2}+J^{3}, & [J^{1},R_{0}]=0,\\

[J^{1},R_{1}]=R_{0}, & [J^{1},T_{1}]=0, &  [J^{2},J^{3}]=0, &  [J^{2},J^{4}]=J^{4},\\

[J^{2},R_{0}]=0 &  [J^{2},R_{1}]=R_{1}, & [J^{2},T_{1}]=-T_{1},& [J^{3},J^{4}]=0,\\

 [J^{3},R_{0}]=-R_{0},& [J^{3},R_{1}]=-R_{1}, & [J^{3},T_{1}]=T_{1}, &  [J^{4},R_{0}]=-R_{1},\\
 
[J^{4},R_{1}]=0, & [R_{0},R_{1}]=0, &   [R_{0},T_{1}]=J^{1}, & [R_{1},T_{1}]=J^{2}-J^{3}.
 
\end{array}\label{KLA}
\end{equation}
We observe that the commutation relation $[J^{4},T_{1}]$ involves terms depending on the generators $T_{k}$ (see \cite{tre09}). However, if we restrict to the value $k=1$ (and $N=0$), the Lie algebra $ \g^{(k)}$ is isomorphic to $\mathfrak{sl}(3)$ and we obtain that the subalgebra $\widehat{g}^{1}$ spanned by $\{J^{1},J^{2},J^{3},T_{1},R_{0},R_{1}\}$, is finite-dimensional. The system thus coincides with the Smorodinsky--Winternitz system \cite{Fris}.

\medskip
\noindent The exactly solvable integrable system given by the Hamiltonian and constants of the motion 
\begin{gather}
 h_{1}=-4 t \partial_{t}^{2}-8 u \partial_{tu}^{2}-4 u \partial_{u}^{2}+4(\omega t-1-a-b)\partial_{t}+(4\omega u-2(2b+1))\partial_{u},\label{Eb1}\\ 
 x_{1}=-4 u (t-u)\partial_{u}^{2}-4 ((b+\frac{1}{2})t-(a+b+1)u)\partial_{u},\label{Eb2}\\ 
 y_{1}=4( (t-u)\partial_{t}^{2}+( \omega (u-t) +a+\frac{1}{2})\partial_{t}.\label{Eb3} 
\end{gather}
and  satisfying the following operator algebra relations
\begin{gather}
 [x_{1},y_{1}]=z_{1},\nonumber\\
 [x_{1},z_{1}]=16 x_{1} y_{1}+2 h_{1}x_{1}-8 z_{1} - 4 \omega (a-b) x_{1}  + 16(-1 +a^{2}+2 ab +b^{2})y_{1} + 2( -1-a +2 a^{2}+b+2ab) \label{tax}\\
[y_{1},z_{1}]=-8 y_{1}^{2}-2 h_{1}y_{1}-\omega^{2} x_{1}
+ 4\omega (a-b) y_{1} +\frac{1}{2}(1+2a) h_{1} \nonumber
\end{gather}
can be expressed algebraically in terms of the generators of the subalgebra $\widehat{g}^{1}$ as follows: 
\begin{gather}
H_{1}=-4 J^{2} J^{1} - 8 J^{3} J^{1}- 4 R_{0}J^{3} + 4 \omega J^{2} - 4 ( (a+b)-1)J^{1} + 4 \omega J^{3} - 2 (2 b+1) R_{0},\label{Ea1}\\
 P_{1}=-4 J^{3} R_{1}+4 J^{3} J^{3} - 4 (b+\frac{1}{2})R_{1}+4 (a+b) J^{3} ,\label{Ea2}\\
 Q_{1}=4 ( J^{2} J^{1}- T_{1}J^{1} + \omega T_{1}-\omega J^{2}+(a+\frac{1}{2})J^{1}.\label{Ea3} 
\end{gather}
In the enveloping algebra $\mathcal{U}(\widehat{g}^{1})$ we have the commutators 
\begin{gather*}
\left[H_1,P_1\right]= 8 R_{0}J^{2}+16 b R_{0}J^{2}+8 R_{1}J^{1}-16 b R_{1}J^{1} +16 R_{0}J^{3}J^{3}-16 R_{1}J^{3}J^{1},\\ 
\left[H_1,Q_1\right]= 4 \omega J^{3}J^{1}- 4 \omega T_{1} R_{0}-4 J^{3}J^{1}J^{1}+4 T_{1}R_{0}J^{1}.
 \end{gather*}
We observe that the polynomials  $P_{1}$ and $Q_{1}$ do not commute with $H_{1}$ at the level of the enveloping algebra, but only as the result of considering the realization (\ref{real}).  Thus the (infinite-dimensional) quadratic algebra determined by (\ref{tax}) 
 is only valid for the given realization, and is not an algebraic consequence of the underlying hidden symmetry algebra.

\medskip
\noindent In this situation, we inspect whether there exist higher-order operators (elements) in the enveloping algebra $\mathcal{U}(\widehat{g}^{1})$ such that they commute with the Hamiltonian as given in (\ref{Ea1}), but only using the commutation relations (\ref{KLA}), i.e., without invoking the explicit realization (\ref{real}). Using symbolic computation packages,\footnote{Specifically, the  NCAlgebra package for \small{MATHEMATICA}$^{\copyright}$.} the following second- and third-order operators that commute with the Hamiltonian (\ref{Ea1}) were found: 

\begin{eqnarray}
A_{1}=\frac{1}{2}(1+2b)J^{1} +\frac{1}{2}(1+2b)R_{0}-J^{3} \omega - \omega T_{1}+ J^{1} J^{3} + J^{1} T_{1} + J^{3} R_{0} + T_{1} R_{0},\label{op1}\\
 B_{1} = \frac{1}{2} (-1 + 2 a) J_1 - J_2 \omega + T_1 \omega 
 + J_1 J_2 + J_1 J_3 -  J_1 T_1 - T_1 R_0, \label{op2}
 \end{eqnarray}
and
\begin{eqnarray}
C_{1}= \frac{1}{4}(-1-2a -2b-4ab)J^{1}+ (1+a+b)J^{3} \omega -\frac{1}{2} T_{1}(\omega -2 a \omega) -\frac{1}{2} R_{1} (\omega + 2b \omega) \nonumber\\
- (1+a+b)J^{1}J^{3}+\frac{1}{2}(1-2a)J^{1}T_{1}+\frac{1}{2}(1+2b)J^{2}R_{0} +\omega J^{3}J^{3}-\omega J^{3} R_{1} +\omega J^{3} T_{1}
\nonumber\\
+ T_{1} R_{0}  -\omega T_{1} R_{1}-J^{1}J^{3}J^{3}-J^{1}J^{3}T_{1} +J^{2}J^{3}R_{0}+J^{2}T_{1}R_{0}. 
\end{eqnarray}
These are genuine  non vanishing differential operators. For the particular realization (\ref{real}), we recover the commutation with the Hamiltonian (\ref{Eb1}). As can somehow be expected, the operators $A_{1}$,  $B_{1}$ and $h_{1}$ are related algebraically, by means of 
\begin{equation*}
A_{1} + B_{1} +\frac{1}{4}  h_{1} =0,
\end{equation*}
implying that, in addition to the Hamiltonian, only one of the found operators can be considered as an (algebraically independent and in fact linearly independent) quadratic integral. 

\medskip
\noindent The crucial step in this approach is to show that the operators $A_{1}$ and $C_{1}$ generate a finite-dimensional polynomial algebra, taking only into account the commutation relations of the generators  $\{J^{1},J^{2},J^{3},T_{1},R_{0},R_{1}\}$, and that the result is completely independent on the particular realization of these generators. Starting from $A_{1}$ and $C_{1}$, the following 
commutators are found: 
\begin{gather*}
 [A_{1},B_{1}]=0,\quad 
[A_{1},C_{1}]=D_{1},\quad 
 [A_{1},D_{1}]=\omega D_{1},\\
 [C_{1},D_{1}]=\frac{1}{2}\{B_{1},D_{1}\}-\frac{1}{2}\{A_{1},D_{1}\}+\frac{1}{2}(a-b)\{A_{1},B_{1}\}\\
+-\frac{1}{2}(1+2a)(a-b)\omega^2 A_1 -\frac{1}{2}(a-b)(1+2b)\omega^2 B_1 +\omega^2 (b-a) \omega^2 C_1\\
 [B_{1},C_{1}]=-D_{1}, 
\end{gather*} 
where $\left\{\circ,\circ\right\}$ denotes here the anticommutator.  

\medskip
\noindent Following this argumentation, we can analyse the finitely-generated sets of elements having at most $p^{th}$-order in the enveloping algebra $\mathcal{U}(\widehat{g}^{1})$ that commute with the Hamiltonian. At the fourth order, three such polynomials $L_1,L_2$ and $L_3$ can be found, the explicit expressions of which are given in Appendix A. In terms of  $A_1,B_1$ and $D_1$, the operators $L_i$ satisfy the following algebraic relations: 

\begin{gather*}
L_3   + \frac{1}{2}  A_1^2  + \frac{1}{2} B_1^2  + \omega (2 a+ b -1)  A_1   +\omega (a-1) B_1 =0,\\
L_2   -\frac{1}{2} A_1^2  - \omega(1+b)  A_1 =0,\\
L_1  -\frac{1}{2} A_1^2  -\frac{1}{2} B_1^2  -2 A_1 B_1  + \frac{1}{2} ( 2 a - 2 b-1) \omega A_1+\omega(1-a)B_1 -  D_1=0.
\end{gather*}
 
 \medskip
 \noindent  It can be shown that at  the fifth order there is another operator $K_1$ that commutes with the Hamiltonian, the explicit expression of which is omitted because of its length. As a polynomial in the operators $A_1,B_1$ and $D_1$ it can be expressed as 
 \begin{equation*}
\begin{split}
 K_1 &+   \frac{1}{\omega}  C_1 A_1    - \frac{1 - a}{\omega}    A_1^2  -\frac{3 + 2 b}{\omega} A_1  B_1  - \frac{1 + b}{\omega}  B_1^2 - \frac{1}{2} (1 + 2 b)   C_1- \frac{5 +13 b}{2} + b^2\\
 & - \frac{1}{2} a (3 + 2 b) A_1  - \frac{1}{2} (-3 + 4 a + 4 a b + 4 b^2) B_1=0.
  \end{split}
\end{equation*} 

\medskip
\noindent The reexamination of the exactly solvable systems in \cite{tre09} shows how the existence of a finite-dimensional polynomial algebra can be derived using only the enveloping algebra of $\mathcal{U}(\widehat{g}^{1})$.

\subsection{Formally equivalent Hamiltonian}
 
Let us now consider another realization of the same underlying $\mathcal{U}(\widehat{g}^{1})$-Lie algebra derived from the coadjoint representation

\begin{equation}
X_i= \sum_{a,b=1}^{n}=x_b c_{ia}^{b} \frac{\partial}{\partial x_a},\quad 1 \leq i \leq n
\end{equation}

The generators are realized by the (linear) operators 

\begin{gather}
J_1= x_1 \partial_{x_2} + x_4 \partial_{x_5},\nonumber\\
J_2= -x_1 \partial_{x_1} + x_5 \partial_{x_5} - x_6 \partial_{x_6}\nonumber\\
J_3=-x_4 \partial_{x_4} - x_5 \partial_{x_5} + x_6 \partial_{x_6}\nonumber\\
R_0= x_4 \partial_{x_3} + x_1 \partial_{x_6}\nonumber\\
R_1 = -x_4 \partial_{x_1} - x_5 \partial_{x_2} + x_5 \partial_{x_3} + (x_2 -x_3) \partial_{x_6}\nonumber\\
T_1=x_6 \partial_{x_2} - x_6 \partial_{x_3} -x_1 \partial_{x_4}- ( x_2 -x_3) \partial_{x_5}\nonumber
\end{gather}

The differential operator corresponding to the Hamiltonian takes the form 
\begin{equation}
\begin{split}
h=& -2( 1 + 2 b ) x_1 \partial_6 - 4 x_1 x_6 \partial_6^2 - 4(-1+a+b) x_4 \partial_5 + 4 x_1 x_5 \partial_5 \partial_6 - 4 x_4 x_6 \partial_5 \partial_6 + 4 x_4 x_5 \partial_5^2 - 4 \omega x_4 \partial_4\nonumber\\
& + 4 x_1 x_4 \partial_4 \partial_6 + 8 x_4^2 \partial_4 \partial_5 + 8 x_4^2 \partial_4 \partial_5  + 2 x_4 \partial_3 - 4 b x_4 \partial_3- 4 x_4 x_6 \partial_3 \partial_6 + 4 x_4 x_5 \partial_3 \partial_5  + 4 x_4^2 \partial_3 \partial_4- 4 a x_1 \partial_2\\
& - 4 b x_1 \partial_2- 4 x_1 x_6 \partial_2 \partial_6+4 x_1 x_5 \partial_2 \partial_5 + 8 x_1 x_4 \partial_2 \partial_4 -4 \omega x_1 \partial_1 + 4 x_1 x_4 \partial_1 \partial_5 + 4 x_1^2 \partial_1 \partial_2.\nonumber
\end{split}
\end{equation}
It can be easily verified that the algebraic polynomials $A_1$, $B_1$, $C_1$ and $D_1$ of degrees 2,2,3 and 4, respectively,  take the form of genuine differential operators of the same order. Therefore, the quadratic algebra is also satisfied, showing that this Hamiltonian (defined on a six-dimensional space) is characterized by the same underlying hidden Lie algebra, integrals of motion and quadratic symmetry algebra.

\subsection{The case $ k=2$}

\noindent For this value of the parameter $k$, the Hamiltonian is given by

\begin{equation}
h_2=-4J^{2} J^{1} -8 J^{3}J^{1}-8 R_1 J^{3} + 4 J^{2} -4( 2(a+b)-1) J^{1}+4 \omega J^{3} - 8 (2 b +1) R_1.\label{cal}
\end{equation}
Up to degree five, there are four polynomials $A_i$ of degrees 2,4,4 and 5, respectively, that commute with the Hamiltonian (\ref{cal}),
satisfying the algebraic relations ( see the Appendix for their explicit expression) 
\begin{gather*}
A_1=-\frac{1}{4}h_2, \nonumber\\
 A_2-\frac{1}{4}A_1^2 - \frac{\omega}{4}(3+ 4a + 4b) A_1 =0
 \end{gather*}
We can close an algebra with $A_3$, $A_4$ and their commutator $A_5=[A_3,A_4]$. 
Redefining the generators as $A'_3 = A_3  +2 A_2   + 2 a \omega A_1$ and $A'_4 = A_4 - (\frac{4}{3} (-2 + a + 3 b)) A_3  - (\frac{1}{3} (-41 + 16 a + 48 b)) A_2   - (\frac{4}{3} (-7 + 5 a + 9 b) \omega)  A_1$, in limit ($\omega = a=b=0 $ ) the following cubic relations are obtained:
\begin{equation}
[A^{\prime}_3,A^{\prime}_5 ]=0,\quad 
[ A^{\prime}_4,A^{\prime}_5]= - \{A^{\prime}_3,A^{\prime}_5\} + 2 A_1^2 A^{\prime}_3  -4 A^{\prime 3}_3.
\end{equation}

\section{ Other Calogero cases }

The last case is in fact related to the $BC_2$ Calogero models \cite{tu98}. There are other cases that have been studied from the point of view of exact solvability and related to certain realizations of $\mathbb{gl}(3)$. Realizations with matrix differential operators have been described in \cite{yu13}. We consider the basis $\left\{E_{ij},T_k^{\pm}\right\}$ with commutators 

\begin{equation}
\begin{array}[c]{lll}
[E_{ij},E_{kl}]=\delta_{jk} E_{il} - \delta_{il} E_{kj}, &  [E_{ij},T_{k}^{-}]=-\delta_{ik} T_j^{-}, & [E_{ij},T_{k}^{+}]=\delta_{jk} T_{i}^{+}, \\
 \left[ T_i^{+},T_{j}^{-}\right]=E_{ii}-\delta_{ij} E_{0}, & [E_0,T_i^{+}]=-T_i^{+} , & [E_0.T_i^{-}]=T_i^{-} .\\
 \end{array}\label{CALO}
\end{equation}
The tree-body $A_2$ Calogero Hamiltonian is then expressed as
\begin{equation*}
h_{cA_2}=-2E_{11} T_1^{-} -6 E_{22} T_1^{-} + \frac{2}{3} E_{12}E_{12} -4\omega E_{11} -2(1+3\nu)T_1^{-} -6\omega E_{22}.\label{calo2}
\end{equation*}
In contrast to the description in  \cite{yu13}, that used an explicit differential operator realization in terms of matrix differential operators, we derive a unified set of integrals algebraically, i.e., independently on the particular realization chosen. Up to the cubic order, there are three polynomials that commute with the Hamiltonian  (\ref{calo2}), and given by 
\begin{gather*}
A_1 = 2 b E_{11} - T_{1}^{-} - \frac{1}{3} E_{12} E_{12} + T_1^{-}  E_{11} + 
   3 T_1^{-} E_{22},\nonumber\\
A_2 = E_{11} + 2 E_{22} + E_{11} E_{11} + E_{11} E_{22} + E_{12} E_{21} + 
   E_{22} E_{22} + T_1^{-} T_1^{+} + T_2^{-} T_2^{+},\nonumber\\
A_3 = -2 E_{11} - 2 E_{22} - 2 E_{11} E_{11} - 3 E_{11} E_{22} - 
   E_{22}  E_{22} - 2 T_1^{-} T_1^{+} - 2 T_2^{-} T_2^{+} - 
   E_{11} E_{11} E_{22}\nonumber\\
   + E_{11} E_{12} E_{21} - E_{11} E_{22} E_{22} + E_{12} E_{21} E_{22} + 
   T_1^{-} E_{12} T_2^{+} - T_1^{-} E_{22} T_1^{+} - 
   T_2^{-} E_{11} T_2^{+} + T_2^{-} E_{21} T_1^{+}
 \end{gather*}   
It can be verified that there are no additional independent operators up to order five. The commutators the Abelian algebra
\begin{equation*}
\left[ A_i,A_j\right]=0,\quad 1\leq i<j\leq 3.
\end{equation*}

\medskip
\noindent 
Similar findings are observed for the case of the three-body Sutherland model with $A_2$ Weyl group, having the Hamiltonian 
\begin{equation*}
\begin{split}
h_{sA2}=&  -2 E_{11} T_1^{-} - 6 E_{22} T_1^{-} + \frac{2}{3} E_{12} E_{12} -2 (1+ 3\nu) T_1^{-} + \frac{\alpha^4}{24} E_{21} E_{21} \\
&  - \frac{\alpha^{2}}{6} ( 3 E_{11} E_{11} + 8 E_{11} E_{22} + 3 E_{22} E_{22} + ( 1 + 12 \nu ) (E_{11} + E_{22} ))
 \end{split}
\end{equation*}
and commuting polynomials up to order three
\begin{equation*}
\begin{split}
A_1=&  \frac{1}{12} (a_2^2 + 12 a_1 a_2^2) E_{11} + \frac{1}{12} (a_2^2 + 12 a_1 a_2^2) E_{22} + 
 3 a_1 T_1^{-} + \frac{1}{4} a_2^2 E_{11} E_{11} + \frac{2}{3} a_2^2 E_{11} E_{22}\\
 &  -\frac{1}{3} E_{12} E_{12} - \frac{1}{48} a_2^4 E_{21} E_{21} + \frac{1}{4} a_2^2 E_{22} E_{22} + 
 T_1^{-} E_{11} + 3 T_1^{-} E_{22},\\
  A_2 =& E_{11} + 2 E_{22} + E_{11} E_{11} + E_{11} E_{22} + E_{12} E_{21} + 
 E_{22} E_{22} + T_1^{-} T_1^{+} + T_2^{-} T_2^{+},\\
A_3 =& 2 E_{22} - E_{11} E_{22} + 2 E_{12} E_{21} + E_{22} E_{22} - 
  E_{11} E_{11} E_{22} + E_{11} E_{12} E_{21} - E_{11} E_{22} E_{22}  \\
 & + E_{12} E_{21} E_{22} + T_1^{-} E_{12} T_{2}^{+} - 
  T_1^{-} E_{22} T_1^{+} - T_2^{-} E_{11} T_2^{+} + T_2^{-}  E_{21} T_1^{+}.
 \end{split}
\end{equation*}
As before, up to order 5 there is no other independent polynomial, as all fourth- and fifth-order polynomial is algebraically dependent. This again leads to the Abelian algebra
\begin{equation*}
\left[ A_i,A_j\right]=0,\quad 1\leq i<j\leq 3.
\end{equation*}

\section{Higher-order polynomial algebras deduced from $\mathfrak{gl}(3)$}

\noindent In a slightly different manner, the exact solvability of superintegrable systems was discussed in the paper \cite{tem01}  using  the reductive Lie algebra $\mathfrak{gl}(3,\mathbb{R})$ as hidden symmetry algebra. The authors considered the following realization:
\begin{gather}
J_{1}= \partial_{t},\quad J_{2}=\partial_{u} ,\quad J_{3}=t \partial_{t},\quad  
 J_{4}=u\partial_{u} ,\quad J_{5}=u\partial_{t},\quad J_{6}=t\partial_{u} \nonumber\\
  J_{7}=t^{2} \partial_{t}+t u \partial_{u} - n t ,\quad J_{8}=t u \partial_{t} + u^{2}\partial_{u} - n u,\quad X=n. \label{su3}
 \end{gather}

\noindent In terms of these generators, the commutation relations are given by 

\begin{equation}
\begin{array}[c]{llll}
[J_{1},J_{2}]=0, &   [J_{1},J_{3}]=J_{1}, &  [J_{1},J_{4}]=0, &  [ J_{1},J_{5}]=0,\nonumber \\
\left[J_{1},J_{6}\right]=J_{2}, &  [J_{1},J_{7}]=2J_{3}+J_{4}-n, &  [J_{1},J_{8}]=J_{5}, &  [J_{2},J_{3}]=0,\nonumber\\
\left[J_{2},J_{4}\right]=J_{2},  & [J_{2},J_{5}]=J_{1}, & [J_{2},J_{6}]=0, &  [J_{2},J_{7}]=J_{6},\nonumber\\
\left[J_{2},J_{8}\right]=J_{3}+2 J_{4}-n ,&  [J_{3},J_{4}]=0, &  [J_{3},J_{5}]=-J_{5}, & [J_{3},J_{6}]=J_{6},\nonumber\\
\left[ J_{3},J_{7}\right]=J_{7}, &  [J_{3},J_{8}]=0, &  [J_{4},J_{5}]=J_{5}, &  [J_{4},J_{6}]=-J_{6},\nonumber\\
\left[J_{4},J_{7}\right]=0, &  [J_{4},J_{8}]=J_{8}, &  [J_{5},J_{6}]=-J_{3}+J_{4}, &  [J_{5},J_{7}]=J_{8}\nonumber\\
\left [J_{5},J_{8}\right]=0, & [J_{6},J_{7}]=0, & [J_{6},J_{8}]=J_{7}, & [J_{7},J_{8}] =0.\\
 \end{array}\label{KLB}
\end{equation}
The maximal parabolic subalgebra is generated by $\{J_1,J_2,J_3,J_4,J_5,J_6\}$.
 The main feature of the preceding realization lies in the fact that the four classical (and quantum) superintegrable systems on the real plane admitting two independent first integrals quadratic in the momenta, in addition to the Hamiltonian, can be rewritten in terms of the maximal parabolic subalgebra of $\mathfrak{sl}(3,\mathbb{R})$ (see \cite{tem01} for details). From these four superintegrable systems, we consider the two first, as the analysis of the remaining cases essentialy reduces to that of Case I. 

\subsection{Case I}

\noindent We consider the following differential operators

\begin{eqnarray}
h^{I}=-2 t \partial_{t}^{2}-2 u \partial_{u}^{2}+ 2t\partial_{t}+2u \partial_{u}-(2 p_{1}+1)\partial_{t} -(2 p_{2}+1)\partial_{u} +1 +p_{1}+p_{2},\nonumber\\
x_{c}^{I}=2t\partial_{t}^{2}-2 u \partial_{u}^{2}-2 t \partial_{t}+2u\partial_{u}+(2p_{1}+1)\partial_{t}-(2p_{2}+1)\partial_{u}-p_{1}+p_{2}, \label{op21}\\
x_{r}^{I}=4 t u (\partial_{t}-\partial_{u})^{2} +2 ((2p_{1}+1)u-(2p_{2}+1)t)(\partial_{t}-\partial_{u})-(p_{1}+p_{2})^{2},\nonumber
\end{eqnarray}
where the first operator one represents the Hamiltonian of the system and the remaining ones the quadratic first integrals:  
\[
[h^{I},x_{c}^{I}]=[h^{I},x_{R}^{I}]=0.
\]
\noindent As differential operators, the generators $x_{c}^{I},x_{r}^{I}$  span the following algebra:
\begin{gather}
 [x_{c}^{I},x_{r}^{I}]=x_{cr}^{I},\nonumber\\
\left[ x_{r}^{I},x_{cr}^{I}\right]=16 x_{c}^{I}x_{r}^{I}-8 x_{cr}^{I} -16 (-1 +p_{1}^{2}+p_{1}p_{2}+p_{2})x_{c}^{I} 
-16 (p_{1} -p_{2})x_{r}^{I} -16 (-p_{1}+p_{1}^{2}+p_{2}-p_{2}^{2})h^{I},\\
\left[x_{c}^{I},x_{cr}^{I}\right]=-8 (x_{c}^{I})^{2}+8 (h^{I})^{2} + 16 (p_{1}-p_{2})x_{c}^{I} 
 +16 x_{r}^{I} +16 (1+p_{1}+p_{2}) h^{I}.\nonumber
\end{gather}
In terms of the maximal parabolic subalgebra of $\mathfrak{gl}(3 )$, the differential operators (\ref{op21}) can be expressed as 

\begin{gather}
H^{I}=-2 J_{3}J_{1}-2 J_{4}J_{2}+2J_{3}+2J_{4}-(2p_{1}+1)J_{1}-(2p_{2}+1)J_{2},\nonumber\\
P_{c}^{I}=2J_{3}J_{1}-2J_{4}J_{2}-2J_{3}+2J_{4}+(2p_{1}+1)J_{1}-(2p_{2}+1)J_{2},\label{op22}\\
P_{R}^{I}=4J_{3}J_{5}+4 J_{4}J_{6}-8 J_{3}J_{4}+ 2(2p_{1}+1)J_{5}
-2 ( 2p_{2}+1)J_{3} -2( 2p_{1}+1)J_{4}+2(2p_{2}+1)J_{6}, \nonumber
\end{gather}
from which we obtain the relations $ [H^{I},P_{c}^{I}]=0$ and 
\begin{equation*}
\begin{split}
 [H^{I},P_{R}^{I}]=&  4 J_{3} J_{2} + 8 p_{2} J_{3}J_{2}+4 J_{4}J_{1} + 8 p_{1}J_{4}J_{1}+4 J_{5}J_{2}-8 p_{1}J_{5}J_{2} + 4 J_{6}J_{1}-8 p_{2}J_{6}J_{1}\\
 & +8 J_{4}J_{3}J_{1}+8 J_{4}J_{3}J_{2}-8 J_{5}J_{3}J_{2}-8 J_{6}J_{4}J_{1}.
 \end{split}
\end{equation*}

\noindent The interesting fact is that the operator $P_{R}^{I}$, seen as a polynomial in the generators of the subalgebra of $\mathfrak{gl}(3 )$, does not commute with the Hamiltonian as given in  (\ref{op22}). The third order polynomial on right-hand side of $[H^{I},P_{R}^{I}]$  only vanishes when the differential operator realization given by (\ref{op21}) is used. This points out that the preceding quadratic algebra is only satisfied at the operator level,  and not as a polynomial in the enveloping algebra of $\mathfrak{gl}(3 )$. 
 
\smallskip
\noindent Like before, we search systematically for all second- and third-order polynomials in the generator of $\mathfrak{gl}(3 )$ that commute with the Hamiltonian as given in (\ref{op22}). In this case, the following eight polynomials can be found: 
\begin{equation*}
\begin{split}
A_{1}= &  -\frac{ 2 - n + (4  - 2 n) p_1}{2}J_5 - 
  \frac{2 - n - (4  - 2 n) p_1}{2} J_4  
- \frac{2 - n + (4 - 2 n) p_2}{2} J_6  - 
  \frac{2 - n - (4 - 2 n) p_2}{2} J_3 \\
&  - J_3^2 + (3 - 2 n) J_3  J_4 + ( n-2) J_3 J_5 
	+ ( n-2) J_4 J_6 - J_1 J_4  J_7 +  J_1  J_6  J_8- J_2 J_3 J_8 + J_2 J_5 J_7+  J_3^2 J_4 \\
	&  + J_3 J_4^2 - J_3 J_5 J_6 -   J_4 J_5 J_6,\\
	A_2= & - \frac{2p_1+1}{2} J_5 + \frac{2p_2-1}{2}J_3+ \frac{1-2p_2}{2}  J_2 J_3+  \frac{2p_1+1}{2}  J_2 J_5+ J_3 J_4 - J_3 J_5 - 
  J_2 J_3 J_4 + J_2 J_3 J_5,\\
  A_3= & \frac{2p_1-1}{2} J_4  - \frac{1+2p_2}{2}J_6 + \frac{1 - 2 p_1}{2}  J_1 J_4+ \frac{1 + 2 p_2}{2}  J_1 J_6+ J_3 J_4 - J_4 J_6 -  J_1 J_3 J_4 + J_1 J_4 J_6,\\
  A_4= & \frac{2p_1-1}{2} J_4  -  \frac{1+2p_1}{2}J_5 + \frac{1+2p_2}{2} J_3  -  \frac{1+2p_2}{2} J_6+ J_3 J_4 - J_3 J_5 - J_4 J_6 + J_5 J_6,\\
  A_5 =& (n-1)J_4    - \frac{1+2p_1}{2} J_5 + \frac{2p_2-1}{2} J_3   +  \frac{2p_1-1}{2} J_1 J_4+ \frac{1 - 2 p_2}{2} J_2 J_3+ J_2 J_8 - J_3 J_5-J_4^2+ J_1  J_3 J_4 - J_2 J_3 J_4,\\
  A_6= &-J_4 + \frac{2p_2-1}{2} J_2  + J_2 J_4,\\
  A_7= &  (n-1)J_3  + \frac{2p_1-1}{2} J_4   -  \frac{1+2p_2}{2} J_6+  \frac{1 - 2 p_1}{2}  J_1 J_4+ J_1 J_7+ \frac{2p_2-1}{2}  J_2 J_3
  - J_3^2  - J_4 J_6 - J_1 J_3 J_4 + J_2 J_3 J_4,\\
  A_{8}= & -J_3 + \frac{2p_1-1}{2} J_1 + J_1 J_3. 
 \end{split}
\end{equation*}
 
\noindent Some of these operators can be related with the original generators $P_{c}^{I}$, as the following identities hold: 
 \begin{equation*}
A_{6}-\frac{1}{4}H^{I}+\frac{1}{4}P_{c}^{I}=0,\quad A_{8}-\frac{1}{4}H^{I}-\frac{1}{4}P_{c}^{I}=0.
\end{equation*}
We observe that the polynomials $\{A_2,A_3,A_4,A_6,A_8\}$ all belong to the enveloping algebra of the maximal parabolic subalgebra of $\mathfrak{gl}(3)$.

\noindent It should be observed that for the realization (\ref{su3}) as differential operators, both  $A_1$, $A_{5}$ and $A_{7}$ are expressible as linear combinations of $\left\{A_2,A_3,A_4,A_6,A_8\right\}$, hence they are described in terms of the maximal parabolic subalgebra : 
\begin{equation*} 
 A_1 - (2 -n)  A_4  + 2 n=0,\quad A_{5}-A_{2}+A_{8}+n =0,\quad A_{7}+A_{2}-A_{4}-A_{8}+n =0.
\end{equation*}
 
This illustrates again the differences between the enveloping algebra of Lie algebra and that of their realization as differential operators, and that in the purely algebraic setting, other polynomials involving all generators play a role in closing a quadratic algebra.

\medskip
\noindent Prior to determine other algebraic relations between the generators, we find the minimal polynomial algebra $\frak{A}$ that $\left\{A_1,\dots ,A_8\right\}$ generate, and analyze whether the complete set of polynomials that commute with the Hamiltonian is finitely generated and coincides with $\frak{A}$.  Considering the commutators of the polynomials $\left\{A_1,\dots ,A_8\right\}$, we further consider the change of basis given by 
\begin{equation*}
A_2\rightarrow A_3+A_2,\quad  A_6\rightarrow A_6 + A_8
\end{equation*}
which ensures that we obtain a quadratic algebra:
\begin{equation}
\begin{split}
 [A_{1},A_{3}]=  & A_4 A_6 (-2+n)-2 A_4 A_8 (-2+n)+A_6 (2-n-4 p_1+2 n p_1)+A_3 (-6+3 n+2 p_1-n p_1+2 p_2-n p_2)+\nonumber\\
 &A_4 (-2 p_1+n p_1+2 p_2-n p_2)-2 A_8 (2-n-2 p_1+n p_1-2 p_2+n p_2)+\frac{1}{2} A_2 (6-3 n-4 p_2+2 n p_2)    ,\nonumber\\
    [A_{1},A_{8}]= & A_2 (-2+n)-2 A_3 (-2+n)+\frac{1}{2} A_6 (-2+n+4 p_1-2 n p_1)+A_8 (2-n-2 p_1+n p_1-2 p_2+n p_2)  ,\nonumber\\ 
     [A_{2},A_{3}]=& -A_4 A_6+2 A_4 A_8+A_6 (1-2 p_1)+\frac{1}{2} A_2 (3-2 p_2)+A_4 (-p_1+p_2)+A_3 (-3+p_1+p_2)+2 A_8 (-1+p_1+p_2) ,\nonumber\\ 
      [A_{2},A_{8}]=& -A_2+2 A_3+\frac{1}{2} A_6 (-1+2 p_1)+A_8 (1-p_1-p_2),\nonumber\\
      [A_{3},A_{4}]= & A_4 A_6-2 A_4 A_8+A_6 (-1+2 p_1)+A_3 (3-p_1-p_2)+A_4 (p_1-p_2)-2 A_8 (-1+p_1+p_2)+\frac{1}{2} A_2 (-3+2 p_2),\nonumber\\
       [A_{3},A_{8}]= & A_3-A_6 A_8+A_8^2+A_8 (\frac{1}{2}-p_1)+\frac{1}{2} A_6 (-1+2 p_1),\nonumber\\
        [A_{5},A_{8}]= & -A_2+A_3+A_6 A_8-A_8^2+\frac{1}{2} A_8 (1-2 p_2),\nonumber\\
         [A_{7},A_{8}]=& A_3-A_6 A_8+A_8^2+A_8 (\frac{1}{2}-p_1)+\frac{1}{2} A_6 (-1+2 p_1) 
\end{split}
\end{equation}

\noindent The generators $\left\{A_2,A_3,A_4,A_6,A_8\right\}$ are related through a cubic constraint: 
\begin{equation*}
\begin{split}
A_2 A_3 - A_2 A_6 - A_3 A_8 +  \frac{2p_1+1}{2} A_4 A_6   + 
 \frac{2p_1-1}{2} A_6^2 -  \frac{2p_1-1}{2}A_3 A_6  + \frac{2p_2-1}{2}A_4 A_8\\
    -  \frac{2 p_1 - 2 p_2 - 4 p_1 p_2+1}{4} A_2   - 
  \frac{2 p_1 + 2 p_2 - 4 p_1 p_2-5}{4} A_3  - \frac{3 - 6 p_1 - 2 p_2 + 4 p_1 p_2}{4} A_6 \\
 - \frac{2p_2-1}{2}A_2 A_8 +\frac{2p_2-1}{2} A_8^2 +  \frac{2 p_1 - 2 p_2 - 4 p_1 p_2+1}{4} A_4 +  \frac{2 p_1 - 2 p_2 - 4 p_1 p_2+1}{4}  A_8   - A_4 A_6 A_8  =0
\end{split}
\end{equation*}

We obtain a quadratic algebra of dimension 8. We observe that those polynomials belonging to the enveloping algebra of maximal parabolic subalgebra $\{A_2,A_3,A_4,A_6,A_8\}$ of $\mathfrak{gl}(3)$ give rise to a five-dimensional quadratic subalgebra.

\subsection{Case II}

\noindent The second type given in \cite{tem01} possesses the following Hamiltonian and first integrals:

\begin{equation}
\begin{split}
h^{II}=& -\partial_{t}^{2}-2 u \partial_{u}^{2}+2 t \partial_{t}
 + ( 2u -1 -2 p_2)\partial_{u} +\frac{3}{2}+p_2, \\ 
x_{c}^{II}=& 2 \partial_{t}^{2}-4 u \partial_{u}^{2}-4 t \partial_{t}
+2(2 u -1 -2 p_2) \partial_{u}-1+2p_2,\label{ope31}\\
x_{p}^{II}= &-4 t u \partial_{u}^{2} + 4 u \partial_{tu}^{2}
-(2 u-1-2 p_2) \partial_{t} - 2 t (1+2 p_2)\partial_{u}. 
\end{split}
\end{equation}

\noindent Besides the relations $[h^{II},x_{c}^{II}]=0,\quad   [h^{II},x_{p}^{II}]=0$, these differential operators satisfy the following commutation relations: 
\begin{gather*}
   [x_{c}^{II},x_{p}]=x_{cp}^{II},\quad [x_{c}^{II},x_{cp}^{II}]=64 x_{p}^{II},\\
\quad  [x_{p}^{II},x_{cp}^{II}]=-6 x_{c}^{II}-8 h^{II} x_{c}^{II}+8 (h^{II})^{2} + 8 ( -3 -4 p_{2}+4 p_{2}^{2} ),
\end{gather*}
where 
\begin{gather*}
x_{cp}^{II}=-32u \partial^{3}_{tuu}-16(1+2p_2-2u)\partial^{2}_{tu}+32tu\partial^{2}_{uu}+8(1+2p_2-2u)\partial_t+16(1+2p_2)t\partial_u.
\end{gather*}
 
\noindent Expressed in terms of the generators of the subalgebra $\frak{A}$, we have that 
\begin{gather}
H^{II}=-J_{1}J_{1} -2 J_{4}J_{2}+2 J_{3}+2 J_{4}-(1+2p_2)J_{2}+\frac{3}{2}+p_2,\nonumber\\
P_{c}^{II}=2 J_{1}J_{1} -4 J_{4}J_{2}-4 J_{3}+4 J_{4}
-(1+2 p_2) J_{2}-1+2p_2,\label{op32}\\
P_{p}^{II}=-4 J_{4}J_{6} +4 J_{1}J_{4}-2 J_{5}
 +(1+2p_2)J_{1}-(2+4p_2)J_{6}.\nonumber
\end{gather}
\noindent While the operator $P_{c}^{II}$ still commutes with the Hamiltonian as a polynomial in the $J_i$'s, $P_{p}^{II}$ leads to a quadratic element: 
\begin{equation*}
 [H^{II},P_{c}^{II}]=0,\quad [P^{II},P_{p}^{II}]=4 (J_{5}J_{2}- J_{4}J_{1}).
\end{equation*}
The latter commutator only vanishes when evaluating the generators $J_i$ at the realization (\ref{ope31}). We conclude that the quadratic algebra is only valid as operator algebra and is not a consequence of the underlying Lie algebra but 
of the explicit choice of realization as differential operator.

\medskip
\noindent Analyzing the enveloping algebra in search of second- and third-order polynomials that commute with the Hamiltonian in (\ref{op32}) at the purely algebraic level, we again find eight such polynomials, given respectively by 
   
\begin{equation*}
\begin{split}
A_{1}= & (n-2)J_3  - J_3^2 + (1 - n) J_3  J_4 + (n-2 ) J_5 J_6- J_1 J_4 J_7 + J_1 J_6 J_8- J_2 J_3 J_8\\
& +  J_2 J_5 J_7+ J_3^2J_4 + J_3 J_4^2 - J_3 J_5 J_6 - J_4 J_5 J_6,\\
A_{2}=&  -J_3 + \frac{1 + 2 p_2}{2} J_1 J_6 + \frac{1 - 2 p_2}{2}  J_2 J_3+  J_3 J_4 - J_5 J_6-\frac{1}{2} J_1^2 J_4+\frac{1}{2} J_1 J_2 J_5\\
& + J_1 J_4 J_6 - J_2 J_3 J_4,\\
A_{3}=& -  (1 + 2 p_2)J_6  + \frac{2 p_2-1}{2}   J_1 J_2- 2 J_4 J_6 + J_1 J_2 J_4,\\
A_{4}=& (n-1)J_4  + J_2 J_8 - J_3 J_4 - J_4^2+\frac{1}{2} J_1^2J_4 -\frac{1}{2} J_1 J_2 J_5,\\
A_{5}=& -J_5 + \frac{2p_2-1}{2} J_1  -  (1 + 2 p_2)J_6+ J_1 J_4 +   J_2 J_5 - 2 J_4 J_6,\\
A_{6}=& -J_4 + \frac{2 p_2-1}{2} J_2   + J_2 J_4,\\
A_{7}=& (n-2)J_3 + J_1 J_7 - J_3^2- J_5 J_6 -   \frac{1}{2} J_1^2J_4+\frac{1}{2} J_1 J_2 J_5,\\
A_{8}=& -2 J_3 + J_1^2.
\end{split}
\end{equation*}

\noindent The polynomials $\{A_2,A_3,A_4,A_5,A_6,A_8\}$  all belong to the enveloping algebra of the maximal parabolic subalgebra of $\mathfrak{gl}(3)$. We observe that the generators $A_{1}$, $A_4$ and $A_{7}$ reduce to linear combinations of $\left\{A_1,A_2,A_3,A_5,A_6,A_8\right\}$ when evaluated at the realization 
(\ref{ope31})
\begin{equation*}
A_1-2n=0,\quad A_{8}-2 A_{7}-2n =0,\quad A_{8}+2 A_{4}+2n =0.
\end{equation*}
This clearly illustrates how the problem of finding integrals in the realization and as abstract polynomials are different and need to be looked at separately. These polynomials span a 8-dimensional quadratic algebra, the commutators of which are given by 
\begin{equation*}
\begin{split}
 [A_{3},A_{5}]= & -2 A_2-2 A_6^2+2 A_6 A_8+\frac{1-2 p_2}{2} A_8 + (1+2 p_2)A_6,\\
  [A_{3}, A_{6}] =&  -A_3,\\
	[A_{3},  A_{8}]=&  2 A_3,\\
		[ A_{5}, A_{6}]=&  -2 A_3+A_5,\\
   [A_{5},  A_{8}]= & 4 A_3-2 A_5,
\end{split}
\end{equation*} 
All the other commutators vanish. The generators $\left\{A_1,A_2,A_3,A_5,A_6,A_8\right\}$ are moreover related by means of a cubic polynomial  
\begin{equation*}
\frac{1}{2} A_3^2 - \frac{1}{2} A_3 A_5 + A_2 A_6 - A_6^2 -  \frac{1}{2}A_6^2 A_8 +   \frac{2p_2-1}{2} A_6   +  \frac{2p_2+1}{4} A_6 A_8  -\frac{2p_2+1}{2} A_2 =0.
\end{equation*}
Again, the commutators involving only the polynomials related to the maximal parabolic subalgebra of $\mathfrak{gl}(3)$ detetermine a 6-dimensional quadratic subalgebra.

\subsection{Formally equivalent Hamiltonian}

\noindent Using the coadjoint representation, we obtain another realization in eight variables 
\begin{gather*}
J_1= x_1 \partial_3 + x_2 \partial_6 + ( 2 x_3 + x_4) \partial_7 + x_5 \partial_8,\\
J_2= x_2 \partial_4 + x_1 \partial_5 + x_6 \partial_7 + ( x_3 + 2 x_4 ) \partial_8,\\
J_3=- x_1 \partial_1 - x_5 \partial_5 + x_6 \partial_6 + x_7 \partial_7,\\ 
J_4 =-x_2 \partial_2 + x_5 \partial_5 - x_6 \partial_6 + x_8 \partial_8,\\
J_5 -x_1 \partial_2 + x_5 \partial_3 -x_5 \partial_4 + ( -x_3 +x_4) \partial_6 +x_8 \partial_7,\\
J_6=-x_2 \partial_1 - x_6 \partial_3 + x_6 \partial_4 -(-x_3 +x_4)\partial_5 +x_7 \partial_8,\\
J_7=-(2x_3 +x_4) \partial_1 - x_6 \partial_2 - x_7 \partial_3 -x_8 \partial_5,\\
J_8 =-x_5 \partial_1 -(x_3 + 2x_4 ) \partial_2 - x_8 \partial_4 - x_7 \partial_6.
\end{gather*}
In this realization, we obtain the two following Hamiltonians, both of them possessing first integrals compatible with the quadratic algebra: 
\begin{equation}
\begin{split}
 h_I =&  -(x_3+2p_2 x_3 +2 x_4 + 4 p_2 x_4 -x_5 +2 p_1 x_5 -2 x_8 ) \partial_8  - 2(x_3 +2 x_4 )x_8 \partial_8^2-2 x_3 \partial_7 - 4 p_1 x_3 \partial_7 \nonumber\\
 & - x_4 \partial_7- 2p_1 x_4 \partial_7 + x_6 \partial_7 + 2p_2 x_6 \partial_7 + 2 x_7 \partial_7- 2 x_5 x_7 \partial_5 \partial_7 - 2 x_6 x_8 \partial_7 \partial_8 - 4 x_3 x_7 \partial_7^2 - 2 x_4 x_7 \partial_7^2\nonumber\\
 & +  x_2 \partial_6 - 2 p_1 x_2 \partial_6 + 2 x_3 x_6 \partial_6 \partial_8 + 4 x_4 x_6 \partial_6 \partial_8 - 2 x_5 x_6 \partial_6 \partial_8 - 4 x_3 x_6  \partial_6 \partial_7 -2 x_4 x_6 \partial_6 \partial_7 + 2 x_6^2 \partial_6 \partial_7\nonumber\\
 &  - 2 x_2 x_7 \partial_6 \partial_7 - 2 x_2 x_6 \partial_6^2 - x_1 \partial_5 - 2 p_2 x_1 \partial_5 - 2 x_3 x_5 \partial_5 \partial_8  -4 x_4 x_5 \partial_5\partial_8 +2 x_5^2 \partial_5 \partial_8 - 2 x_1 x_8 \partial_5 \partial_8\nonumber\\
 & + 4 x_3 x_5 \partial_5 \partial_7 + 2 x_4 x_5 \partial_5 \partial_7 -2 x_5 x_6 \partial_5 \partial_7 + 2 x_2 x_5 \partial_5 \partial_6 + 2 x_1 x_6 \partial_5 \partial_6 -2 x_1 x_5 \partial_5^2 + x_2 \partial_4-2 x_2 \partial_2 \nonumber\\
 & -2 p_2 x_2 \partial_4 \partial_8 - 2 x_2 x_8 \partial_4 \partial_8 +2 x_2 x_6 \partial_4 \partial_6-2 x_2 x_5 \partial_4 \partial_5 + x_1 \partial_3 -2 p_1 x_1 \partial_3 -2 x_1 x_7 \partial_3 \partial_7- 2 x_1 x_6 \partial_3 \partial_6 \nonumber\\
 & + 2 x_1 x_5 \partial_3 \partial_5+ 2 x_2 x_3 \partial_2 \partial_8 + 4 x_2 x_4 \partial_2 \partial_8 + 2 x_2 x_6 \partial_2 \partial_7 + 2 x_1 x_2  \partial_2 \partial_5 + 2 x_2^2 \partial_2 \partial_4 - 2 x_1 \partial_1 + 2 x_1 x_5 \partial_1 \partial_8\nonumber\\
 &   + 4 x_1 x_3 \partial_1 \partial_7+ 2 x_1 x_4 \partial_1 \partial_7 + 2 x_1 x_2 \partial_1 \partial_6 + 2 x_1^2 \partial_1 \partial_3
\end{split}
\end{equation}

and 

\begin{equation}
\begin{split}
 h_{II} =&  \left( \frac{3}{2} +p_2\right) - ((1+2p_2)(x_3 +2 x_4)-2 x_8) \partial_8 - x_5^2 \partial_8^2 - 4 x_4 x_8 \partial_8^2 + ( -2 x_1 +x_6 -2 p_2 x_6 +2 x_7) \partial_7 \nonumber\\
 & - 4 x_3 x_5 \partial_7 \partial_8  - 2 x_3 x_8 \partial_8^2- 2 x_4 x_5 \partial_7 \partial_8 - 4 x_3^2 - 4 x_3 x_4 \partial_7^2 -x_4^2 \partial_7^2 -2 x_2 x_5 \partial_6 \partial_8+ 2 x_3 x_6 \partial_6 \partial_8\nonumber\\
 &  + 4 x_4 x_6 \partial_6 \partial_8 -4 x_2 x_3 \partial_6 \partial_7 -2 x_2 x_4 \partial_6 \partial_7 +2 x_6^2 \partial_6 \partial_7-x_2^2 \partial_6^2 - (1+2 p_2) x_1 \partial_5 -2 x_3 x_5 \partial_5 \partial_8\nonumber\\
 & -4 x_4 x_5 \partial_5 \partial_8 - 2 x_1 x_8 \partial_5 \partial_8 -2 x_5 x_6 \partial_5 \partial_7 + 2 x_1 x_6 \partial_5 \partial_6 -2 x_1 x_5 \partial_5^2 + x_2 \partial_4- 2 p_2 x_2 \partial_4-2 x_1 \partial_1\nonumber\\
 & - 2 x_2 x_8 \partial_4 \partial_8 + 2 x_2 x_6 \partial_4 \partial_6 -2 x_2 x_5 \partial_4 \partial_5 -2 x_1 x_5 \partial_3 \partial_8 -4 x_1 x_3 \partial_3 \partial_7 -2 x_1 x_4 \partial_3 \partial_7 -2 x_1 x_2 \partial_3 \partial_6\nonumber\\
 &  -x_1^2 \partial_3^2 -2 x_2 \partial_2+ 2 x_2 x_3 \partial_2 \partial_8 + 4 x_2 x_4 \partial_2 \partial_8 + 2 x_2 x_6 \partial_2 \partial_7 + 2 x_1 x_2 \partial_2 \partial_5 +2 x_2^2 \partial_2 \partial_4    - 2 x_6 x_8 \partial_7 \partial_8.
\end{split}
\end{equation}
A classification of formally-equivalent models is certainly an open problem, as it depends heavily on the classification of realizations of Lie algebras by vector fields, a problem that has currently only been solved for low-dimensional Lie algebras \cite{Po03}.
 
\section{Generalized algebraic Hamiltonians}

\noindent The preceding examples illustrate that the space $\mathfrak{M}_d$ of polynomials commuting with an algebraic Hamiltonian $H$ can differ considerably from one case to another. In this context, it is reasonable to inspect to which extent the existence of commuting polynomials depends on the relation among the coefficients of the Hamiltonian $H$, and whether by an appropriate variations of these constants a given exactly solvable system can be generalized to a parameterized family with similar properties.

\medskip
\noindent As first example, we consider a generalized version of the (algebraic) Hamiltonian (\ref{Ea1})
\begin{equation}
 H_{1}=a_{1} J^{2} J^{1} + a_{2} J^{3} J^{1}+ a_{3} R_{0}J^{3} + a_{4}  J^{2} + a_{5} J^{1} + a_{6} J^{3} + a_{7} R_{0},
\end{equation}
where the parameters $a_i$ are supossed to be free. Besides this quadratic polynomial in $\mathcal{U}(\widehat{g}^{1})$, which is the Hamiltonian itself, there are no other polynomials of the second, third or fourth order that commute with the Hamiltonian $H_1$, unless certain algebraic relations among the parameters are also imposed. This shows that the generic analysis does not necessarily lead to the construction of quadratic algebras, as additional constraints on the coefficients are required to ensure the existence of algebraic higher-order first integrals. Ultimately, such relations among the coefficients are  heavily dependent  on the underlying hidden symmetry algebra considered.

\medskip
\noindent Let us now consider the following parameterized generalization of the Hamiltonian in (\ref{op22}):
\begin{equation}
H'^{I}=a_{1} J_{3}J_{1} + a_{2}  J_{4}J_{2}+ a_{3} J_{3}+ a_{4} J_{4}+ a_{5} J_{1}+a_{6} J_{2}.
\end{equation}
As before, no relation between the parameters is assumed. In contrast to the preceding case, here there exist two (independent) polynomials of order 2, 4 of order 3 and 7 of order four that commute with $H'^{I}$. Among these polynomials, those of order two commute with each other and with any of the polynomials of orders three and four. Considering the commutators of the non-quadratic polynomials, additional higher-order operators are obtained, so that higher-order non-Abelian polynomial algebras may exist for this generic case.

\medskip
\noindent Finally, the Hamiltonian   
\begin{equation}
H'^{I}=-2 J_{3}J_{1} + -2 2 J_{4}J_{2}+ 2 J_{3}+ 2 J_{4}- (2 p_{1}+1) J_{1}- (2 p_{2}+1) J_{2} + p_{3} J_{1} J_{2}  + p_{4} J_{1}^{2} + p_{5} J_{2}^{2}
\end{equation}
can be considered as a deformation of the Hamiltonian in (\ref{op22}) by a quadratic term parameterized by $p_3,p_4,p_5$. In this case, two second-order polynomials and one third-order polynomials which commute with the Hamiltonian and with each other can be found, thus leading to an Abelian algebra. The analysis of fourth-order polynomials shows that the three solutions are obtained as commutators of the lower order operators. 

\subsection{A generic system with fixed cubic integral}

\noindent We illustrate by an example how the hidden symmetry algebra of a known exactly solvable system can be used to derive new or alternative systems depending on some parameters, starting with a given first integral. To this extent, let us consider again the subalgebra $\widehat{g}^{1}$ of  $\g^{(1)}$ 
spanned by $\{J^{1},J^{2},J^{3},T_{1},R_{0},R_{1}\}$. Instead of using the Hamiltonian  (\ref{Ea1}), we consider the fixed cubic polynomial 
given by 
\begin{equation*}
Q_2=-J^2+T_1-J^2J^3+J^2R_1-T_1R_1+J^2J^3R_1.\label{pox1}
\end{equation*}
A routine computation shows that the most general quadratic element in  $\mathcal{U}\left(\widehat{g}^{1}\right)$ that commutes with $Q_2$ has the form $a_1P_1+\dots a_4P_4$, where $a_i\in\mathbb{R}$ and 
\begin{equation*}
P_1=J^2-J^2J^3+T_1R_1,\quad P_2=J^1+R_0+J^3R_0,\quad P_3=J^2+J^3,\quad P_4=P_3^2.
\end{equation*}
As algebraic Hamiltonian, we consider a generic nonvanishing linear combination $H=a_1P_1+a_2P_2+a_3 P_3+a_4P_4$. Leaving the coefficients free, the only quadratic polynomials that commute with $H$ are the Hamiltonian itself, as well as the polynomial $Q_1=J^3R_1+R_1-J^3$. 
By construction, it follows that $Q_1$ and $Q_2$ do not commute, and indeed we obtain the fourth-order operator
\begin{equation*}
\left[Q_1,Q_2\right]=-J^2 + J^3  + T_1 - J^3 J^2  + (J^3)^2 + R_1 J^3  - T_1 R_1 +  R_1 J^3 J^2 - 2 R_1 (J^3)^2- (R_1)^2 J^3+ (R_1)^2(J^3)^2.
\end{equation*}
These commutation relations are valid in the enveloping algebra $\mathcal{U}\left(\widehat{g}^{1}\right)$ for any values of the parameters $a_i$, and are moreover independent on any realization of the generators by vector fields. If we now take into account the particular realization (\ref{real}), we obtain the differential operators 
\begin{gather}
h= (a_2+(a_1+a_4)t)\partial_{t}+(a_2+(a_1+a_4)u)\partial_{u}+a_4t^2\partial_{t}^2+2a_4tu\partial_{ut}^2+(a_2+a_4u)u\partial_{u}^2+a_3(t\partial_t +u\partial_u),\nonumber\\
x_1=(t-u)\partial_{u}+(u-t)\partial_{t}+(t^2-2tu)\partial_{tu}^2+tu\partial_{u}^2+t^2u\partial_{uut}^3,\\
y_1=(t-u)\partial_{u}+tu\partial_{u}^2.\nonumber
\end{gather}
We obviously have that $x_1$, $y_1$ are constants of the motion of the Hamiltonian $h$, and that the commutator of $x_1$ and $y_1$
leads to the fourth-order differential operator 
\begin{equation*}
\begin{split}
\left[x_1,y_1\right]= & (t-u)\partial_{t}+(t-u)\partial_{u}+(2tu-t^2)\partial_{tu}^2+(5tu-u^2-2t^2)\partial_{u}^2-t^2u\partial_{uut}^3\\
& +(2tu^2-4t^2u)\partial_{u}^3-t^2u^2\partial_{u}^4. 
\end{split}
\end{equation*} 
We further observe that, while $\left[x_1,\left[x_1,y_1\right]\right]$ provides another operator de order six, as expected, the identity $\left[y_1,\left[x_1,y_1\right]\right]-\left[x_1,y_1\right]=0$ is satisfied. Indeed, the commutator between $Q_1$ and $Q_2$ can be expressed as 
\begin{equation*}
 [Q_1,Q_2]= -Q_1^2  - Q_2  + Q_1.
\end{equation*}
With the redefinition $Q_3=-Q_1^2  - Q_2  + Q_1$, the commutation relations can be formulated as 
\begin{equation*}
 [Q_1,Q_2]=Q_3,\quad [Q_1,Q_3]=Q_1^2  + Q_2 -Q_1,\quad  [Q_2,Q_3]=-2Q_1^3  -\{Q_1,Q_2\}+3 Q_1^2 + Q_2  - Q_1,
\end{equation*}
leading to a cubic algebra \cite{mar09} and such cubic algebra.

\section{Conclusion}
 
\noindent Many approaches have been introduced over the years to generate superintegrable systems and their symmetry algebras. These approaches are based on PDE's,  ladder operators, orthogonal polynomials or reccurence relations, thus they rely on explicit realizations. The hidden algebra approach combines both a purely algebraic construction and explicit realizations of Lie algebras by  differential operators. In this paper we have focused on the algebraic formalism underlying the hidden symmetry method, and more specifically, on the construction of finite-dimensional quadratic polynomial algebras. These algebras are obtained analyzing the subspaces that commute with a Hamiltonian, whenever the latter is written as an element belonging to the enveloping algebra of an appropriate  subalgebra of the hidden symmetry algebra. For three known models studied in \cite{tre09,tem01}, we obtained a $4$-dimensional, a $5$-dimensional, a $6$-dimensional and two $8$-dimensional quadratic algebra, respectively. The symmetry algebra of the Smorodinsky-Winternitz and the $BC_2$-Calogero systems, which correspond to a special cases of the hierarchy studied in \cite{tem01}, can thus be formulated in a purely algebraic way, without referring to an explicit realization by differential operators. This suggests to reexamine other models along the same lines, such as the singular Wolfes model, which is also contained in the family described in \cite{tem01}, and even to see whether the algebraic approach can provide additional information on the higher-order integrals for the parameter values for which the direct analytic approach has turned out to be intractable. 

\medskip
\noindent In principle, there are two possibilities to pursue the analysis. On the one hand, we can analyze deformations or generalizations of a given Hamiltonian and hidden symmetry algebra, in order to find further new or alternative (superintegrable) systems, by suitably characterization of the subspaces $\frak{M}_d$ of polynomials that commute with the Hamitonian $H$.  We presented three examples of such new equivalence in this paper. As such polynomials commute with $H$ in the corresponding enveloping algebra, they will give rise to constants of the motion for any nontrivial realization of the Lie algebra by vector fields. This means that the same operators would lead to essentially different systems for nonequivalent realizations.  Another possibility concerns the classification of classes of polynomial algebras, by means of re-examination of the enveloping algebra of Lie algebras. The first step in this direction would be to establish criteria that allow to decide if for any given (quasi-) exactly solvable system there exist finite-dimensional quadratic polynomial algebras associated to it, and whether the resulting algebras are related to each other, either by extension or some other procedure. Work in these directions is currently in progress.

\section*{Acknowledgements}
IM was supported by the Australian Research Council Future Fellowship FT180100099. RCS was   supported by
the research projects MTM2016-79422-P of the AEI/FEDER (EU) and PID2019-106802GB-I00 /AEI / 10.13039/501100011033
of the Agencia Estatal de Investigaci\'on (Spain).

\section{Appendix A}

\subsection{$k=1$}

\noindent The polynomials $L_1$, $L_2$ and $L_3$ commuting with the Hamiltonian (\ref{Ea1}) are explicitly given by

\begin{equation*}
\begin{split}
L_1 =&   a \omega^2J_3 + \frac{\omega^2}{2} (1 -2 a - 2 b)J_2 + \frac{\omega}{4} (1+ 4 b + 4 b^2)R_0- 2\omega^2 (1- a)T_1 
 - \frac{\omega^2}{2} (1 + 2 b)R_1\\
 & - \frac{\omega}{2}  (-1 + 2 a - 2 a^2  - 2 b  - 2 b^2 ) J_1+ \frac{1}{2} (a^2-a  - b + 2 a b + b^2) J_1^2 + \frac{\omega}{2} (1 - 2 a) J_1 J_3\\
 & +   \frac{1}{2} (1 + a + 3 b + 2 a b + 2 b^2) J_1 R_0+2\omega (1- a) J_1 T_1 +\frac{\omega}{2} ^2 J_2^2+ \omega^2 J_2 J_3
  +   \frac{3 \omega^2}{2} J_3^2\\
  & - \omega^2 J_3 R_1 + \omega^2 J_3 T_1+ \frac{1}{8} (3 + 8 b + 4 b^2) R_0^2+ \frac{\omega}{2} (5  - 2 a) T_1 R_0 -  \omega^2 T_1 R_1 - \omega J_1 J_2^2\\
  & + \frac{1}{2} (2 a + 2 b-1) J_1^2J_2+ (2 a + 2 b-1) J_1^2J_3- 3 \omega J_1 J_2 J_3 + 
   \frac{1}{2} (1 + 2 b) J_1 J_2 R_0 - 3 \omega J_1 J_3^2\\
   &  + (2 + a + 3 b) J_1 J_3 R_0- \omega J_1 J_3 T_1 + \omega J_2 T_1 R_0 - \omega J_3^2R_0 + (1 + b) J_3 R_0^2 +   \frac{1}{2} J_1^2 J_2^2+  2 J_1^2J_3^2\\
   &  + 2 J_1^2 J_2 J_3 + J_1 J_2 J_3 R_0 +   2 J_1 J_3^2 R_0  + \frac{1}{2} J_3^2 R_0^2
\end{split}
\end{equation*}

\begin{equation*}
\begin{split}
L_2 =&   \frac{\omega}{4} (1+ 4 b+ 4 b^2)J_1 +\frac{\omega}{4} (1+ 4 b+ 4 b^2 )R_0 -  \omega^2 (1+ b)J_3- \frac{a\omega^2}{2}  (3  + 2 b)T_1 + \frac{1}{8} (3 + 8 b + 4 b^2) J_1^2 \\
& +  \frac{1}{4} (3 + 8 b + 4 b^2) J_1 R_0+ \omega J_1 T_1 + \frac{1}{2} \omega^2 J_3^2 + \omega^2 J_3 T_1+ \frac{1}{8} (3 + 8 b + 4 b^2) R_0^2+ \omega T_1 R_0 + \frac{1}{2} \omega^2 T_1^2\\
&  + (1 + b) J_1^2J_3 + \frac{1}{2} (1 + 2 b) J_1^2 T_1 - \omega J_1 J_3^2 +   2 (1 + b) J_1 J_3 R_0 - 2 \omega J_1 J_3 T_1 + (1 + 2 b) J_1 T_1 R_0  \\
& - \omega J_1 T_1^2 -  \omega J_3^2 R_0 + (1 + b) J_3 R_0^2 - 2 \omega J_3 T_1 R_0 + \frac{1}{2} (1 + 2 b) T_1 R_0^2-  \omega T_1 ^2R_0 + \frac{1}{2} J_1^2J_3^2 + J_1^2 J_3 T_1\\
&  + \frac{1}{2} J_1^2 T_1^2+ J_1 J_3^2R_0 + 2 J_1 J_3 T_1 R_0  + J_1 T_1^2 R_0 + \frac{1}{2} J_3^2 R_0^2 + J_3 T_1 R_0^2 + \frac{1}{2} T_1^2 R_0^2
 \end{split}
\end{equation*}

 \begin{equation*}
\begin{split}
L_3 =&   -\frac{\omega}{2} (1 - 2 a + 2 a^2 + 2 b + 2 b^2 )J_1 - \frac{\omega}{4} R_0 (1+ 4 b + 4 b^2 )R_0 - \omega^2 (1 - a)J_2+\omega^2 (1+b) J_3\\
&  - \omega^2 ( a - b-3)T_1+  \frac{1}{4} ( 4 a - 2 a^2 - 4 b - 2 b^2-3) J_1^2 +  \frac{\omega}{2} (1 - 2 a) J_1 J_3- \omega J_1 T_1+ \omega^2 J_2 T_1\\
& - \frac{1}{4} (3 + 8 b + 4 b^2) J_1 R_0- \frac{\omega^2}{2} ( J_2 ^2+J_3^2)- \omega^2 J_3 T_1 - \frac{1}{8} (3 + 8 b + 4 b^2) R_0^2
+ \frac{ \omega}{2} (2a-5 ) T_1 R_0\\
& - \omega^2 T_1 ^2+ (1 - a) J_1^2 J_2-(a+b) J_1^2J_3 + (a-b-2) J_1^2T_1 \omega J_1 J_2^2+  \omega J_1 J_2 J_3- 2 \omega J_1 J_2 T_1\\
& + \omega J_1 J_3^2- 2 (1 + b) J_1 J_3 R_0 + \omega J_1 J_3 T_1+ \frac{1}{2} ( 2 a - 4 b-3) J_1 T_1 R_0+   2 \omega J_1 T_1^2 - \omega J_2 T_1 R_0\\
& + \omega J_3^2R_0 -(1 + b) J_3 R_0^2+ 2 \omega J_3 T_1 R_0- \frac{1}{2} (1 + 2 b) T_1 R_0 ^2 + 2 \omega T_1^2 R_0 - \frac{1}{2} J_1^2 J_2^2 -  J_1^2J_2 J_3\\
&+ J_1^2J_2 T_1- J_1^2J_3^2 - J_1^2 T_1^2+  J_1 J_2 T_1 R_0  - J_1 J_3^2 R_0   - J_1 J_3 T_1 R_0 - 2 J_1 T_1^2R_0 -  \frac{1}{2} J_3^2 R_0^2\\
& - J_3 T_1 R_0^2  - T_1^2 R_0^2.
 \end{split}
\end{equation*}

\subsection{Commuting polynomials for  $k=2$}

\noindent The polynomials $A_i$ ($i=1,\dots 4$) up to degree five commuting with the Hamiltonian (\ref{cal}) are given by 

\begin{equation*}
\begin{split}
A_1 =&  (1 + 2 a + 2 b) J_1 + 2 (-1 + 2 b) R_1 - J_2a \omega - J_3 a \omega + J_2 J_1 + 2 J_3 J_1 + 2 R_1 J_3.
 \end{split}
\end{equation*}
\begin{equation*}
\begin{split}
A_2=& \frac{1}{2} (3 - 2 a - 8 b + 4 a b + 4 b^2) R_0 +
  \frac{1}{2} (1 + 4 a  + 4 a^2  + 4 b + 8 a b  + 4 b^2)\omega J_1 - \frac{\omega^2}{4} (3   + 4 a   + 4 b ) J_2  \\
  & + 2   (2 a b + 2 b^2-1- a  + b ) \omega R_1 - \frac{\omega^2}{4} (3  + 4 a   + 4 b )J_3 +  \frac{1}{2} (1 + 3 a + 2 a^2 + 3 b + 4 a b + 2 b^2) J_1^2\\
  & +  \frac{ \omega^2}{4} J_2^2 + \frac{\omega}{2} (1 + 2 a + 2 b ) J_3 J_1+ \frac{1}{2} \omega^2 J_3 J_2 + \frac{\omega^2}{4}  J_3^2 
   + \frac{1}{2} (2b-1) R_0 J_2 + \frac{1}{2} ( 2 a + 6 b--5) R_0 J_3\\
   &  + \frac{1}{2} (1 - 4 a - 6 b + 8 a b + 8 b^2) R_1 J_1 + \omega(1 - 2 b) R_1 J_2 + (3 + 2 a ) \omega R_1 J_3 + (3 - 8 b + 4 b^2) R_1^2+ R_0 J_3 J_2\\
   &  +  \frac{1}{4} (3 + 4 a + 4 b) J_2 J_1^2 - \frac{1}{2} \omega J_2^2 J_1 + \frac{1}{2} (3 + 4 a + 4 b) J_3 J_1`2 - \frac{3}{2} \omega J_3 J_2 J_1 - \frac{1}{2} \omega J_3^2 J_1+ (2b-1 ) R_1 J_2 J_1\\
   & +  R_0 J_3 ^2 + \frac{1}{2} ( 4 a + 12 b-5) R_1 J_3 J_1 - \omega R_1 J_3  J_2 - \omega R_1 J_3^2  + 4 (b-1 ) R_1^2 J_3 +\frac{1}{4} J_2^2 J_1^2 + J_3 J_2 J_1^2+ J_3^2 J_1^2\\
   & + R_1 J_3 J_2 J_1 + 2 R_1 J_3^2 J_1 + R_1^2J_3^2.\\
  \end{split}
\end{equation*}

 \begin{equation*}
\begin{split}
A_3=&   ( 4 a - 4 a^2 + 8 b - 12 a b + 8 a^2 b - 4 b^2 +  8 a b^2-3) R_0+ \frac{\omega^2}{2} T_1 	- 
   2  (1 + 3 a + 2 a^2  + 2 b + 2 a b) \omega J_1\\
   &  -  2  ( 8 a b- 4 a   + 6 b   -3) \omega R_1 + \frac{\omega^2}{2}  (5  + 6 a)J_3 + \frac{\omega^2}{2}  (5 + 4 a)J_2- \frac{1}{2} (5 + 8 a + 4 a^2 + 2 b + 4 a b) J_1^2\\
   &  - \frac{\omega^2}{2}  J_2^2 - (5 + 4 a - 2 b) \omega J_3 J_1 + 2 (1 - 3 a - 3 b + 6 a b + 2 b^2) R_0 J_2- 2 (1 - a - 3 b + 2 a b + 2 b^2) R_1 J_1  \\
   & +2 (3 - 4 a + 2 a^2 - 5 b + 6 a b + 2 b^2) R_0 J_3  + 2 a R_0 T_1+  2 (2b-1)\omega R_1 J_2- 2 (3 + 4 a) \omega R_1 J_3  - 2 \omega R_1  T_1\\
   &  -  2 (3 - 8 b + 4 b^2) R_1^2  - \omega T_1 J_1 - \frac{1}{2} (5 + 4 a) J_2 J_1^2+  \omega J_2^2 J_1 - (2+ 3 a + 2 b) J_3 J_1^2 
    +\omega J_3 J_2 J_1 + \omega J_3 J_3 J_1\\
    & + 2 (2b-1) R_0 J_2^2 + (6 a + 8 b-5) R_0 J_3 J_2 + 4 ( a + b-1) R_0 J_3^2  + R_0 T_1 J_2+R_0 T_1 J_3- 
   2 \omega R_1 J_3 J_2 \\
   &  + (5 - 2 a - 8 b) R_1 J_3 J_1+ 8 (1 - b) R_1^2 J_3 +  R_1 T_1 J_1 + \frac{1}{2} T_1 J_1^2- \frac{1}{2} J_2^2 J_1^2- J_3 J_2 J_1^2
    -\frac{3}{2} J_3^2J_1^2 + 2 R_0 J_3 J_2^2\\ 
    & 3 R_0 J_3^2J_2 + R_0 J_3^3 -    3 R_1 J_3^2 J_1 - 2 R_1^2 J_3^2.
\end{split}
\end{equation*}

 \begin{equation*}
\begin{split}
A_4 =&  -2 (9 - 10 a + 6 a^2 - 4 a^3 - 33 b + 30 a b - 16 a^2 b +  8 a^3 b + 36 b^2 - 20 a b^2 + 8 a^2 b^2 - 12 b^3) R_0\\
& -  (9 + 24 a + 4 a^2  - 16 a^3   + 18 b - 4 a b  -  48 a^2 b  - 12 b^2  - 56 a b^2 - 24 b^3) \omega J_1-  (2 a  + b) \omega^2 T_1 \\
& +\omega^2 (1 - 2 a - 4 b + 4 a b + 4 b^2 )R_2   +  2   (15  + 6 a   - 12 a^2   - 36 b  - 28 a b   + 24 a^2 b   + 
      32 a b^2 + 24 b^3) \omega R_1\\
      & +  (12  + 7 a  - 12 a^2 - 2 b - 20 a b  - 14 b^2) \omega^2 J_3 + \frac{\omega^2 }{2} (25 + 8 a- 16 a^2- 8 b - 32 a b - 24 b^2) \omega^2 J_2 \\
   &   + ( 2 a^2 + 8 a^3 - 6 b + 6 a b + 
      24 a^2 b + 10 b^2 + 28 a b^2 + 12 b^3-10 - 13 a ) J_1^2+ \frac{\omega^2}{2} ( 4 a + 8 b-7 ) \omega^2J_2^2\\
      & + 2 ( 8 a^2  + 5 b  + 8 a b + 6 b^2-8) \omega J_3 J_1 + (1 - 4 a + 8 a^2 - 6 b + 8 a b - 16 a^2 b + 8 b^2) R_0 
  J_2 + \omega^2 R_2 T_1 \\
  &  - 2 (  13 a - 8 a^2 + 4 a^3 + 31 b - 16 a b + 8 a^2 b -  18 b^2-14 ) R_0 J_3+ (1 - 4 a^2 - 2 b) R_0 T_1-  \frac{1}{2} \omega^2 T_1 J_3\\
  & + 2 ( 13 a - 4 a^2 + 27 b - 42 a b + 8 a^2 b - 42 b^2 + 32 a b^2 + 24 b^3-6) R_1 J_1 - 
   \frac{\omega^2}{2} (3+ 2 a- 2 b) J_3^2\\
   &  -  4 \omega (1- a - 5 b + 2 a b + 6 b^2 ) R_1 J_2+ 2\omega( 12 a^2- 4 a   + 20 b+ 12 a b-19) R_1 J_3+ 4 \omega^2 (b-1) J_3 J_2\\
   &  + 4 ( 3 a + 33 b - 8 a b - 36 b^2 + 4 a b^2 + 12 b^3-9) R_1^2+ 2 ( 2 a + 2 b-1) \omega R_1 T_1+ 2 \omega^2 ( a + 2 b-1) R_2 J_3\\
   & + 
   2 (2 b-1) \omega^2 R_2 J_2 + 2 (2 a + b ) \omega T_1 J_1+ \frac{1}{2} (+ 16 a^2 + 8 b + 32 a b + 24 b^2-25 - 8 a ) J_2 J_1^2 - 
   R_0  T_1 J_3\\
   & + (7  - 4 a   - 8 b )\omega  J_2^2J_1 +  \frac{1}{2} (24 a^2 - 8 b + 72 a b + 52 b^2-35 - 34 a ) J_3 J_1^2+ (15   - 4 a   - 16 b )\omega J_3 J_2 J_1\\
   &+ (1 - 2 a - 4 b + 4 a b + 4 b^2) R_0 J_2^2 + (3 - 8 a^2 - 8 b + 8 a b + 8 b^2) R_0 J_3  J_2  -   2 ( a  + 5 b-6) \omega J_3^2 J_1\\
   &-  2 (4 - 2 a + 2 a^2 - 3 b - 2 a b - 2 b^2) R_0 J_3^2 + 4 (1 - 4 b + 4 b^2) R_1 J_2 J_1  - 
   4 ( 2 b-1) \omega) R_1 J_2^2  - \omega^2 J_3^2J_2\\
   &  + (37 - 34 a + 8 a^2 - 104 b + 48 a b + 72 b^2) R_1 J_3 J_1 - 2 (2 a  + 8 b -3)\omega R_1 J_3 J_2 -   4 ( a + 3 b-2) \omega R_1 J_3^2\\
   &  - \frac{1}{2} \omega^2 J_3^3-  4 (3 - 8 b + 4 b^2) R_1^2J_2+ 
   2 (21 - 8 a - 40 b + 8 a b + 20 b^2) R_1^2J_3- 2 ( 2 b-1) R_1^2 T_1 + \omega T_1 J_3 J_1\\
   &   + (1 - 4 a - 4 b) R_1 T_1   J_1 - 2 \omega R_1 T_1 J_2 + 2 \omega^2 R_2 J_3 J_2 + 
   \omega^2 R_2 J_3^2 + (-2 a - b) T_1 J_1^2+ \frac{1}{2} ( 4 a + 8 b-7) J_2^2 J_1^2\\
   & + ( 4 a + 12 b-11) J_3 J_2 J_1^2+ ( 5 a + 13 b-14) J_3^2 J_1^2 + 2 \omega J_3^2J_2 J_1+ \omega J_3^3J_1+ 2 ( 2 b-1) R_0 J_2^3
   - 4 \omega R_1 J_3 J_2^2 \\
   & + 2 ( a + 6 b-3) R_0 J_3 J_2^2 +  4 ( a + 3 b-2) R_0 J_3^2 J_2 + (2 a + 4 b-3) R_0 J_3^2+ R_0 T_1 J_2^2+2 R_0 T_1 J_3 J_2+ R_0 T_1 J_3^2\\
   & + ( 8 a + 20 b-23) R_1 J_3^2J_1 - 2 \omega R_1 J_3^2J_2  -  16 ( b-1) R_1^2 J_3 J_2+  4 ( a + b-1) R_1^2 J_3^2  - 2 R_1^2T_1 J_3 - 2 R_1 T_1 J_3 J_1\\
   & - \frac{1}{2} T_1 J_3 J_1^2 - J_3 J_3 J_2 J_1^2 - \frac{1}{2} J_3 ^3 J_1^2  + 2 R_0 J_3 J_2^3+5 R_0 J_3^2 J_2^2+  4 R_0 J_3^3J_2
   +R_0 J_3^4 - 4 R_1 J_3^2 J_2 J_1- 2 R_1 J_3^3J_1\\
   &  - 4 R_1^2 J_3^2 J_2 -  2 R_1^2 J_3^3.
     \end{split}
\end{equation*}

\end{document}